\documentclass[useAMS,usenatbib]{mn2e}

\usepackage{amssymb}
\usepackage{graphicx}

\newcommand{\sersic}{S\'{e}rsic}
\newcommand{\dirac}[2]{\langle #1 | #2 \rangle}
\newcommand{\eqn}[1]{Equation \ref{eqn:#1}}
\newcommand{\fig}[1]{Figure \ref{fig:#1}}
\newcommand{\sect}[1]{Section \ref{sec:#1}}
\newcommand{\tab}[1]{Table \ref{table:#1}}

\title[Sersiclets]{Sersiclets -- A Matched Filter Extension of Shapelets for Weak Lensing Studies}
\author[W. Ngan et al.]
  {\parbox[]{6.in}{W. Ngan$^1$, L. Van Waerbeke$^1$, A. Mahdavi$^2$, C. Heymans$^{3,1}$, H. Hoekstra$^{2,4}$\\
  \footnotesize $^1$ University of British Columbia, 6224 Agricultural Road,
  Vancouver, BC, Canada V6T 1Z1\\
  $^2$ Department of Physics and Astronomy, University of Victoria, Victoria, BC, Canada V8P 5C2\\
  $^3$ SUPA, Institute for Astronomy, University of Edinburgh, Blackford Hill, Edinburgh, United Kingdom EH9 3HJ\\
  $^4$ Alfred P. Sloan Fellow}}

\begin{document}

\date{\today}

\pagerange{\pageref{firstpage}--\pageref{lastpage}} \pubyear{2009}

\maketitle

\label{firstpage}

\begin{abstract}
The precision study of dark matter using weak lensing by large scale
structure is strongly constrained by the accuracy with which one can
measure galaxy shapes. Several methods have been devised but none
have demonstrated the ability to reach the level of precision
required by future weak lensing surveys. In this paper we
explore new avenues to the existing {\it Shapelets} approach, 
combining a priori knowledge of the galaxy profile with the power of
orthogonal basis function decomposition. This paper discusses the
new issues raised by this matched filter approach and proposes
promising alternatives to shape measurement techniques. In
particular it appears that the use of a matched filter (e.g.
\sersic\ profile) restricted to elliptical radial fitting functions
resolves several well known Shapelet issues.
\end{abstract}



\section{Introduction}

Galaxy shapes provide the unique signature of gravitational lensing
by large scale structure, which has been recognized as a key to the
study of dark matter and dark energy \citep{munshi2008}. A limiting factor is the accuracy with
which one can measure shapes \citep{CH2006,M2007,HJ2008}. Among
the different existing methods, one particularly interesting
approach is the decomposition of galaxy images using basis functions
e.g. \citet{bj2002} or {\it Shapelets} \citep{ref1,rb2003,mr1}. The strengths of this approach rely on
the fact that the shape measurement is analytical and therefore time
efficient as it involves rather small matrix multiplications. {\it
Shapelets} decompose an image into a linear combination of
orthonormal components up to some truncation order, and the shape
parameters are extracted from a least-squares best fit using the 
recomposed (noise free) model. However, the Shapelet type approach suffers
from a few difficulties:

	(i) The choice of the decomposition truncation order is arbitrary.
In 	practice, different lensing groups use radically different
``optimal'' truncation orders. Some prefer low \citep{KK2006},
while others prefer high \citep{BPR2008}, although the $\chi^2$
values for different truncation order could be very different.
Therefore a constant $\chi^2$ criterion to measure the shape does not
appear to be a robust guarantee of unbiased shape measurement.

	(ii) ``Easy cases'' such as large and bright elliptical galaxies are
	poorly fitted. This suggests that a good fit for low
	signal-to-noise galaxies does not necessarily mean that the shape
	has been correctly measured, since it could just be buried in the sky
	noise. This is the overfitting problem.

	(iii) Basis decomposition has too many degrees of freedom for shear
	measurement, since ideally we are only interested in two numbers (or
	six if we include the flexion). This is where galaxy morphology and
	shear measurement are clearly two different problems.

All of those problems have one common origin, namely the choice of
the zeroth order weight function -- Gaussian functions for both
Cartesian \citep{ref1} and Polar Shapelets \citep{mr1}, as well as \citet{bj2002}. 
Unfortunately, Gaussian functions are poor matches to
real galaxy profiles. Ideally, we would like the zeroth order to be
as close as possible to the real profile, and leave to the basis
decomposition the task to fit departures from this ``typical''
profile.

Currently the most promising shape measurement method uses a
bayesian model fitting approach \citep{MKH2007,K2008}. This
method does not suffer from the same issues as Shapelets, but is
limited by the strong galaxy profile prior.

In this paper, we investigate how the change of the weight
function affects the basis decomposition method, and how it leads
naturally to a hybrid method which combines Shapelets and fitting
techniques. We choose to focus on the \sersic\ profile (hence the
term Sersiclets), but our discussion can be extended to any profile\footnote{While
working on the concepts discussed in
this paper, we became aware of similar investigations using
exponential and hyperbolic sech functions (Kuijken \& van Uitert in prep).},
e.g. Moffat profile for ground based point spread function (PSF).
\sect{methodology} introduces the notation and
gives a technical description of the new fitting functions.
\sect{experiment} shows the impact of those fitting functions on
shape fitting and decomposition. Finally, \sect{conclusion} summarizes 
our work so far and future possibilities.

Note that in this paper we choose not to discuss the PSF deconvolution.
Indeed, the problems we mentioned
earlier affect equally the measurement of galaxy shapes whether or
not the galaxies are convolved with a PSF, and Shapelets are a popular approach
because their Gaussian properties allow for very efficient PSF treatment.
The PSF deconvolution
issue goes beyond this work because it depends on how the PSF
is measured and interpolated between stars. Moreover the approach
developed here could as well be applied to the PSF profile
measurement separately, and then used later to address the
deconvolution step through a forward convolution model fitting
method. See for example \citet{K2008}.


\section{Methodology}
\label{sec:methodology}





\subsection{Basis functions in polar coordinates}
\label{sec:basis_functions} In 1D, all polynomials $P_k(x)$ of
degree $k$ are orthonormal with respect to a weight function $w(x)$
if they satisfy
\begin{equation}
	\int_a^b P_i(x) P_j(x) w(x)dx = \delta_{ij}.
\end{equation}
A particular choice of weight function $w(x)$ uniquely determines
the family of polynomials (e.g. for the Cartesian Shapelets, a Gaussian weight
defines the Hermite polynomials). A complete set of polynomials can
be useful for decomposing an arbitrary function $f(x)$ as a linear
combination of {\it basis functions} $\chi_n(x)=P_n(x)[w(x)]^{1/2}$ such that
\begin{equation}
	f(x) = \sum_{n=0}^\infty A_n \chi_n(x).
\end{equation}

In 2D, the basis functions can be
represented using polar coordinates. Following the intuition for
Polar Shapelets, we separate our basis functions
$\chi_{mn}(r,\phi)$ into the radial component $R_n(r)$
and the angular component $e^{im\phi}$. We also assume that the
weight function $w(r)$ has no angular dependence. The 2D basis
functions $\chi_{mn}(r,\phi)$ in polar coordinates would be in the
form
\begin{equation}
	\chi_{mn}(r,\phi)=R_n(r)[w(r)]^{1/2}e^{im\phi}.
	\label{eqn:defRn}
\end{equation}
The orthonormality is then written as
\begin{equation}
	\int_0^a r\ dr \int_0^{2\pi}d\phi\ \chi_{mn}^*(r,\phi)\chi_{m'n'}(r,\phi) =\delta_{mm'}\delta_{nn'}
\end{equation}
where * denotes complex conjugate. The orthonormality requirements for radial and
angular parts, $\int_0^a R_nR_{n'}w(r)rdr=\delta_{nn'}$ and
$\int_0^{2\pi} e^{i(m'-m)\phi}d\phi = 2\pi\delta_{mm'}$, can be
satisfied independently. In particular, $R_n(r)$ is an orthonormal polynomial of
degree $n$ with respect to the weight function $w(r)$. The integration limit $a$
for the radial component will be discussed in \sect{integration_limits}.

With an orthogonal and complete set of basis functions, an arbitrary image 
$f(r,\phi)$ can then be decomposed into
\begin{equation}
	f(r,\phi) \approx \sum_{n=0}^{n_\rmn{max}}\sum_{m=-n}^n A_{mn} \chi_{mn}(r,\phi)
	\label{eqn:linear_combo}
\end{equation}
where the complex basis coefficients $A_{mn}$ satisfy $A_{mn}^*=A_{-mn}$ so that $f(r,\phi)$ is wholly real.
We will refer $n_\rmn{max}$ as ``order'' in the following sections.

Readers familiar
with Polar Shapelets may notice that our radial component here only requires $n$, and $m$ increases in steps of 1.
In Polar Shapelets, the radial component requires both $m$ and $n$, and $m$ increases in steps of 2.
Our choice to completely decouple the radial and the angular components is of mere convenience, which comes with 
the cost that our set of ``basis functions'' is no longer complete. Consequently, our set of
{\it fitting functions} cannot decompose an arbitrary image. As we shall see in \sect{completeness}
and \sect{basis_reduction}, though, the lack of completeness is not a hindrance
to decomposing images of galaxies for weak lensing, as galaxies follow \sersic\ profiles \citep{sersic1968} and
are not arbitrary in general.





\subsection{Weight function}
The basis functions in Shapelets often require high order
polynomials to describe galaxy shapes accurately because galaxies'
radial light profiles do not match the weight functions.
Galaxies' light profiles are well described by \sersic's empirical
formula \citep{peng2002}:
\begin{equation}
	I(r) = I(k)\exp[-b_\lambda(r/k)^{1/\lambda}+b_\lambda]
	\label{eqn:sersic}
\end{equation}
where $k$ is a scale radius, and $\lambda$ is known as the \sersic\ index. For $0.5\lesssim \lambda
\lesssim 10$, $b_\lambda=2\lambda-1/3$. We use a parameterized form
of \eqn{sersic} as our weight function:
\begin{equation}
	w(r) \equiv \exp\left[-(2\lambda-1/3)\left(\frac{r}{k}\right)^{1/\lambda}\right].
	\label{eqn:weightfunction}
\end{equation}



\subsection{Radial component}
\label{sec:radial_component}
The radial component involves a non-trivial
computational step, as $R_n(r)$ must
satisfy the orthonormality requirement described in
\sect{basis_functions}. We obtain $R_n(r)$ by the Gram-Schmidt 
process\footnote{http://mathworld.wolfram.com/Gram-SchmidtOrthonormalization.html}.
Using Dirac notation $\dirac{R_i}{R_j} \equiv \int R_i(r)R_j(r)w(r)rdr$, 
the Gram-Schmidt process generates each $R_n(r)$ by a recurrence relation:
\[
	R_n(r) = \left( r -
\frac{\dirac{rR_{n-1}}{R_{n-1}}}{\dirac{R_{n-1}}{R_{n-1}}}
\right)R_{n-1} -
\frac{\dirac{R_{n-1}}{R_{n-1}}}{\dirac{R_{n-2}}{R_{n-2}}}R_{n-2}
\]
where $R_0(r) \equiv 1$, and $R_1(r)=\left(r-\frac{\dirac{rR_0}{R_0}}{\dirac{R_0}{R_0}}\right)R_0$.
After the recurrence step, each $R_n(r)$ is individually normalized.




\subsection{Integration limits}
\label{sec:integration_limits}

When generating the radial
polynomials (\sect{radial_component}), a sensible integration
limit must be chosen. In Shapelets
one can integrate $r$ from 0 to $\infty$
thanks to the Gaussian function's localized profile. For \sersic\
functions in general, however, the profile may not be localized enough to
allow for an infinitely large domain. This is generally true for any galaxy and
stellar profile used as a weight function.

\begin{figure}
	\centering
	\includegraphics[width=3.4in]{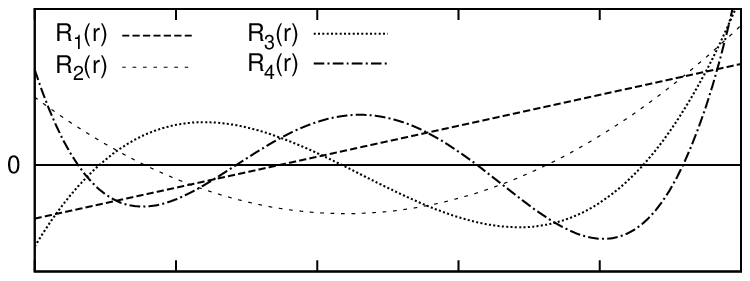}
	\includegraphics[width=3.4in]{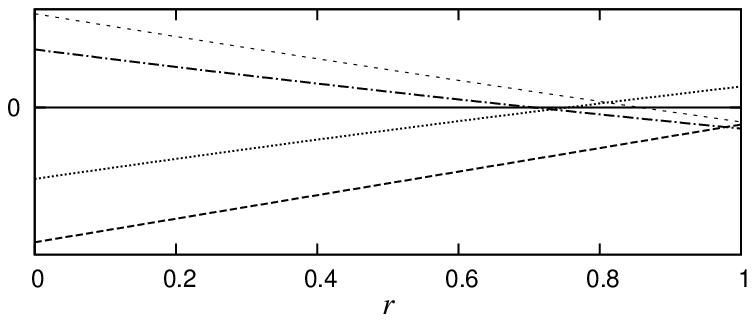}
	\caption{Orthonormal polynomials of degrees 1, 2, 3, and 4 with respect to $w(r)=\exp[-7.67(r/0.5)^{1/4}]$.
 Orthonormality holds in $0<r<1$ (upper panel) and $0<r<\infty$ (lower panel). The functions in the lower panel are
  hardly distinguishable, hence lack linear independence. This is due to large variation in the polynomials
  coefficients, so the high order details are not visible. The functions in the top panel are distinguishably different.}
	\label{fig:infinite_polys}
\end{figure}

The problem with using an infinitely large domain is the lack of mutual independence
among the fitting functions. In order to construct a model as a linear
combination of fitting functions, each fitting function must be distinct so that there is
no redundancy in their shapes. \fig{infinite_polys} shows the polynomials
that are generated using a weight function (\eqn{weightfunction}) with $\lambda=4$.
We find that limiting the orthogonality to a finite domain preserves linear
independency better than extending to an infinitely large domain.

We conveniently choose $0<r<1$ for our domain. For a square image stamp of $2N\times2N$ pixels,
$r$ would be normalized to have units of $1/N$ pixels. It also allows us to constrain
$0< k < 1$, as the scale radius is always positive, and a galaxy should be well captured in a stamp.


\subsection{Completeness}
\label{sec:completeness}
Although the fitting functions of Sersiclets
are indeed mutually orthonormal, they are not necessarily complete. As a result,
\eqn{linear_combo} does not necessarily converge, even at high $n_\rmn{max}$.
The basis functions in both Cartesian and Polar Shapelets are complete, meaning that
\eqn{linear_combo} can converge for any arbitrary $f(r,\phi)$ as $n_\rmn{max} \to \infty$.

\begin{figure}
	\includegraphics[width=3.2in]{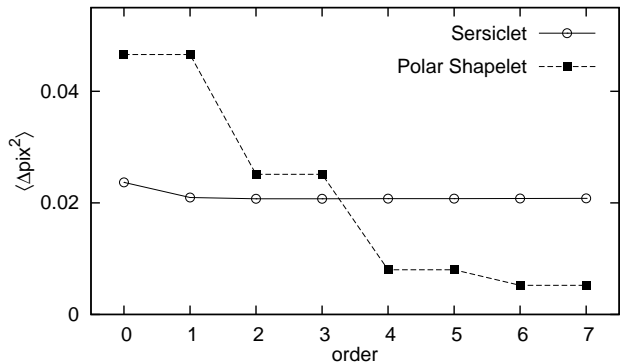}
	\caption{Convergence of \eqn{linear_combo}, represented by the average difference squared
per pixel $\langle \Delta {\rm pix}^2\rangle$ when decomposing a high resolution and noiseless
 elliptical profile. Polar Shapelets are clearly more efficient at describing arbitrary profiles.}
	\label{fig:chi2_convergence}
\end{figure}

The incompleteness of Sersiclets is shown in \fig{chi2_convergence}. It
shows the average difference squared
per pixel $\langle \Delta {\rm pix}^2\rangle$ when decomposing a noiseless elliptical object on a
$128\times128$ grid by integrating \eqn{linear_combo} with $\chi_{mn}^*$, and
exploiting orthonormality to obtain each $A_{mn}$.
The reconstruction using Sersiclets does not
improve even as the order increases; in fact, the reconstruction 
becomes slightly worse because higher order contributions are 
as small as the discretization error. Clearly, Sersiclets fail to decompose even a
simple elliptical object.

For this reason, we would not attempt the decomposition by 
including higher order fitting functions. In the next section we take a different approach --
throwing away all circularly asymmetric components. The lack of
contribution by those components is an important rationale of our technique to 
reduce the set of fitting functions.



\subsection{Fitting set reduction}
\label{sec:basis_reduction}

It is clear from \sect{completeness} that the fitting set should be reduced in
order to take advantage of the matched filter due to the lack
of completeness. In the following we reduce our set of fitting
functions to only the circularly symmetric components ($m=0$),
and we introduce ellipticities by transforming the now perfectly
circular model by ``scaling'' and ``rotating'', which yield unique
values of $e_1$ and $e_2$. This is similar to the process described in
\citet{bj2002}.

The set of ``reduced Sersiclets'' has two advantages; eliminating
the $m\neq0$ components not only cures the overfitting problem, but
it also offers a dramatic increase in speed as the number of terms
in \eqn{linear_combo} now increases like $O(n_\rmn{max})$ rather
than $O(n_\rmn{max}^2)$. It also provides a direct estimate of $e_1$
and $e_2$, which are treated as asymmetric scaling
parameters for the fitting function.

We focus on fitting profiles that are smooth,
centrally peaked, and elliptical in general. These fitting functions are
not suitable for studying galaxy morphology,
as they cannot provide information about a galaxy's detailed
structure. In weak gravitational lensing studies, however, the details
in the typical faint images analyzed are dominated by noise and should not be fitted.
Therefore our reduced fitting set offers a natural regularization process
which is missing in the standard Shapelet approach. Our method is
a hybrid of Shapelets and fitting techniques, where we do
allow some decomposition into fitting functions, but those fitting functions are by construction
axisymmetric and therefore prevent isophote mixing (i.e.
overfitting) as higher order fitting techniques do.



\section{Experiment}
\label{sec:experiment}

\begin{figure}
	\centering
	\begin{tabular}{ccc}
	\vspace{1pt}\\
	\includegraphics[width=0.9in]{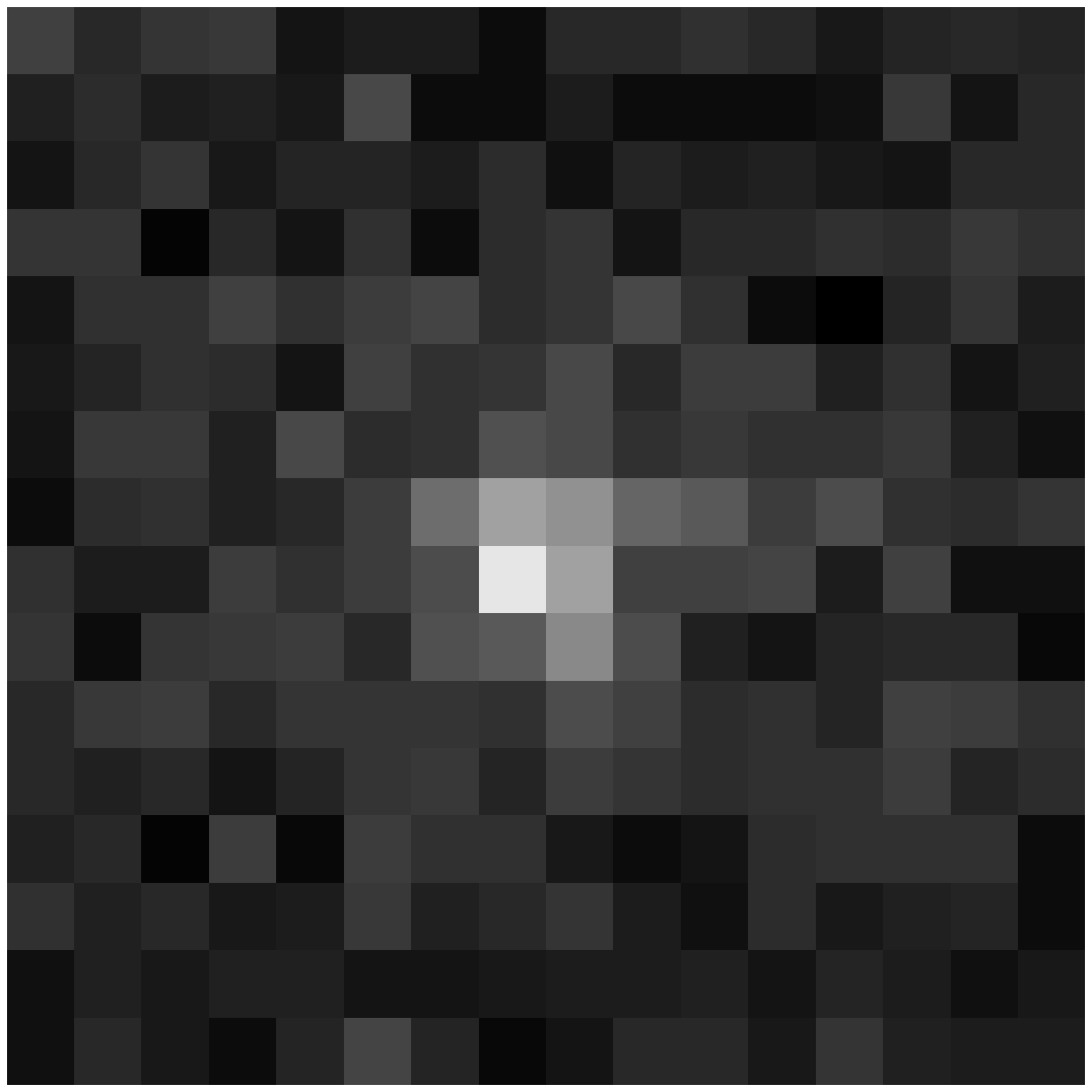} & \includegraphics[width=0.9in]{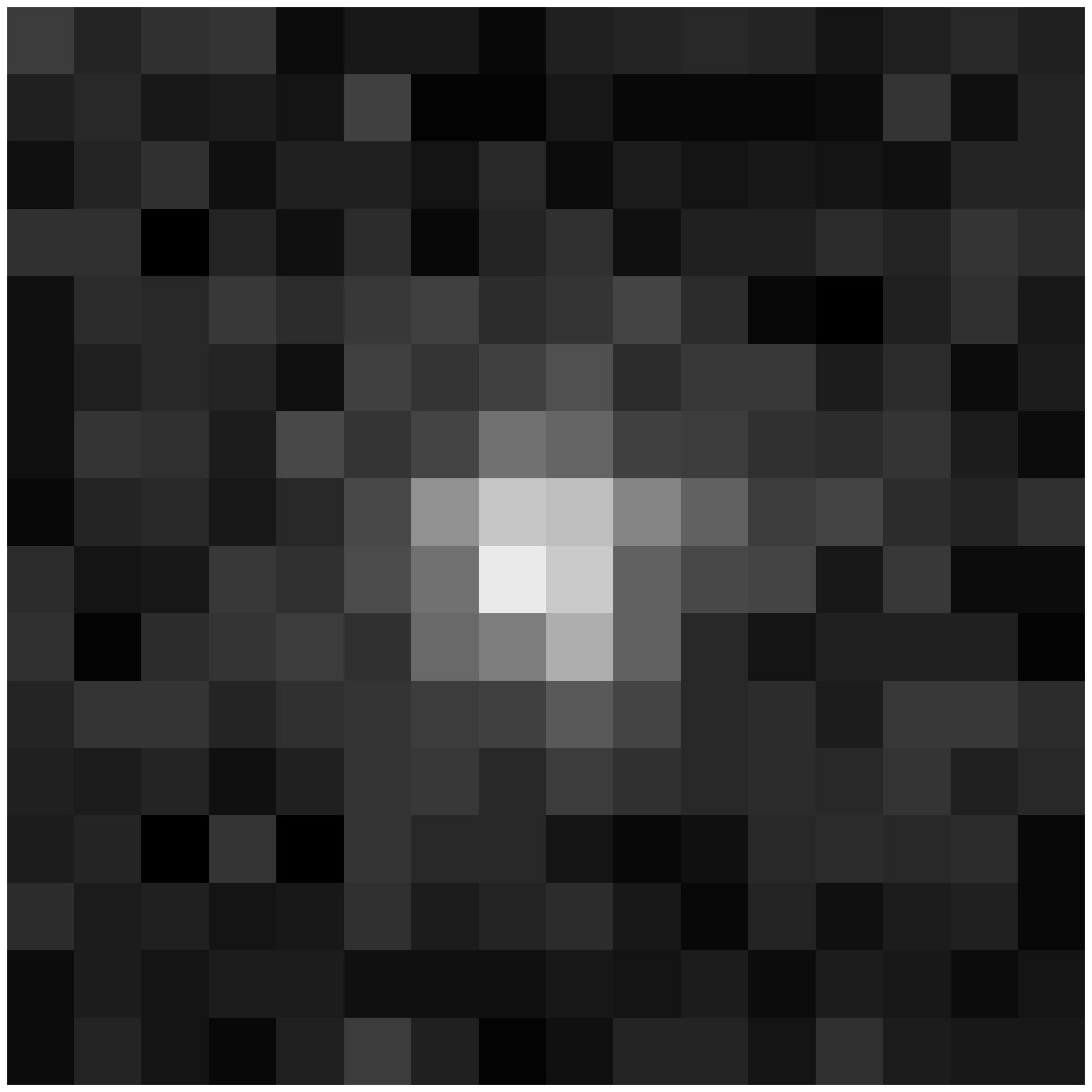} & \includegraphics[width=0.9in]{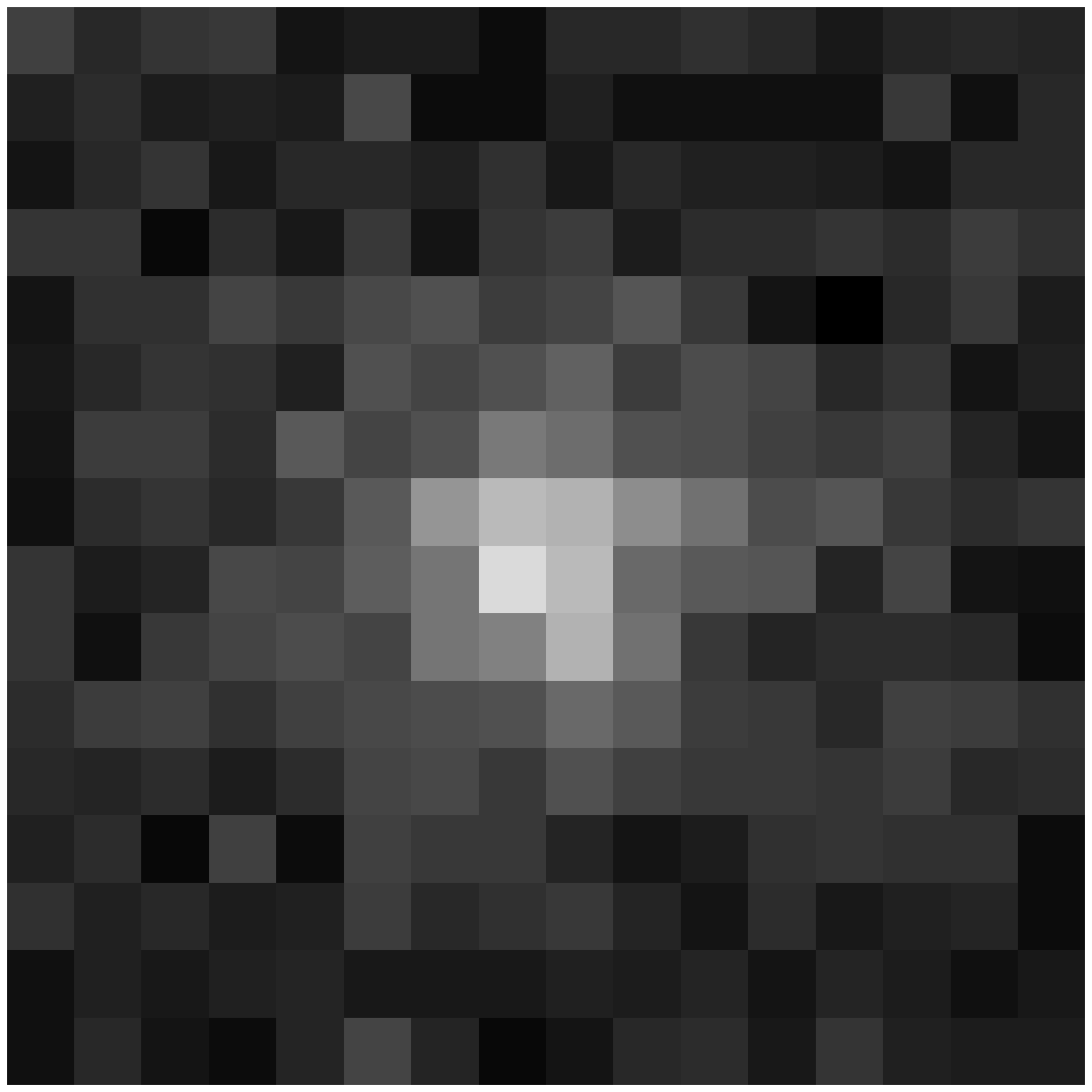}\\
	de Vaucouleurs & exponential & generic \sersic\\\\
	\includegraphics[width=0.9in]{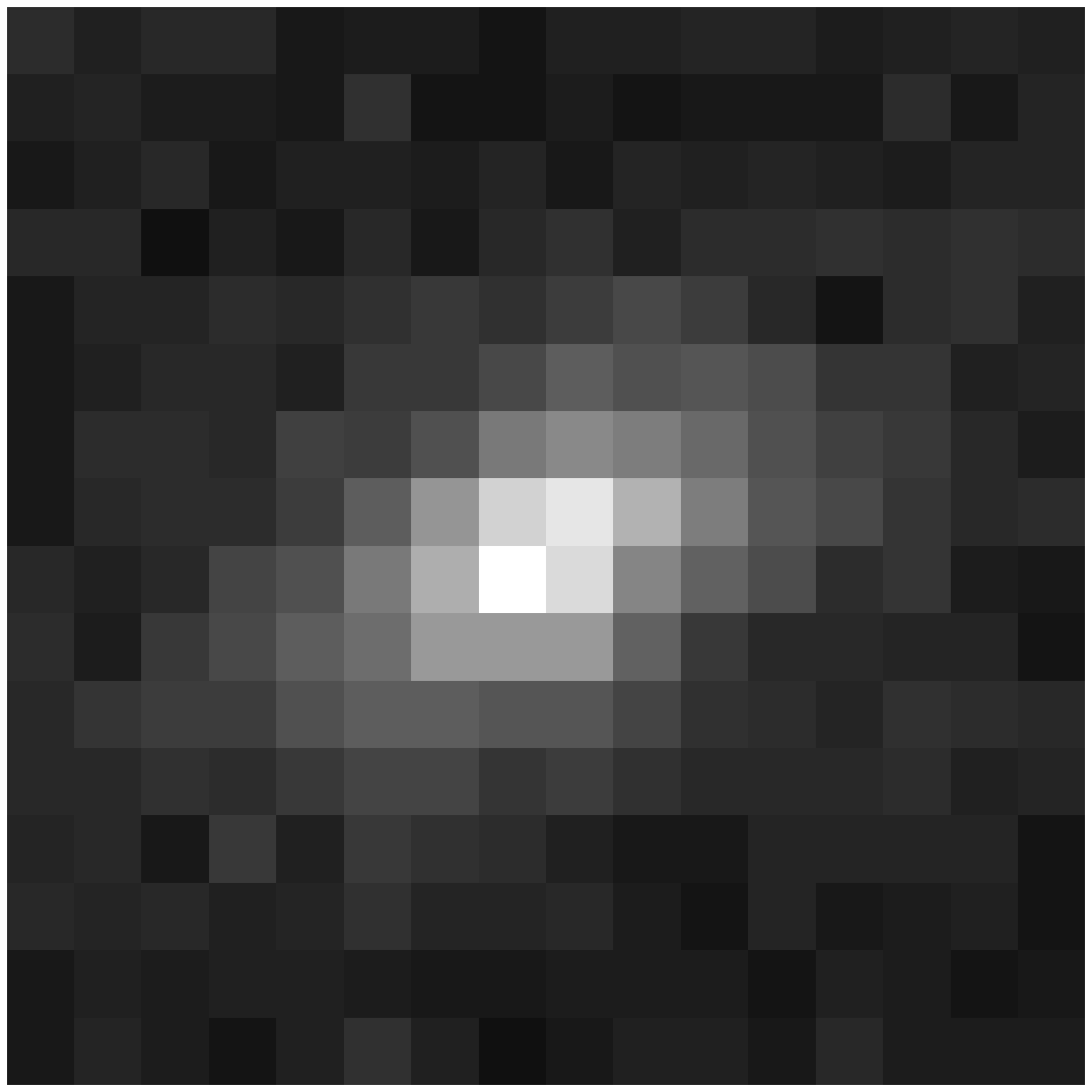} & \includegraphics[width=0.9in]{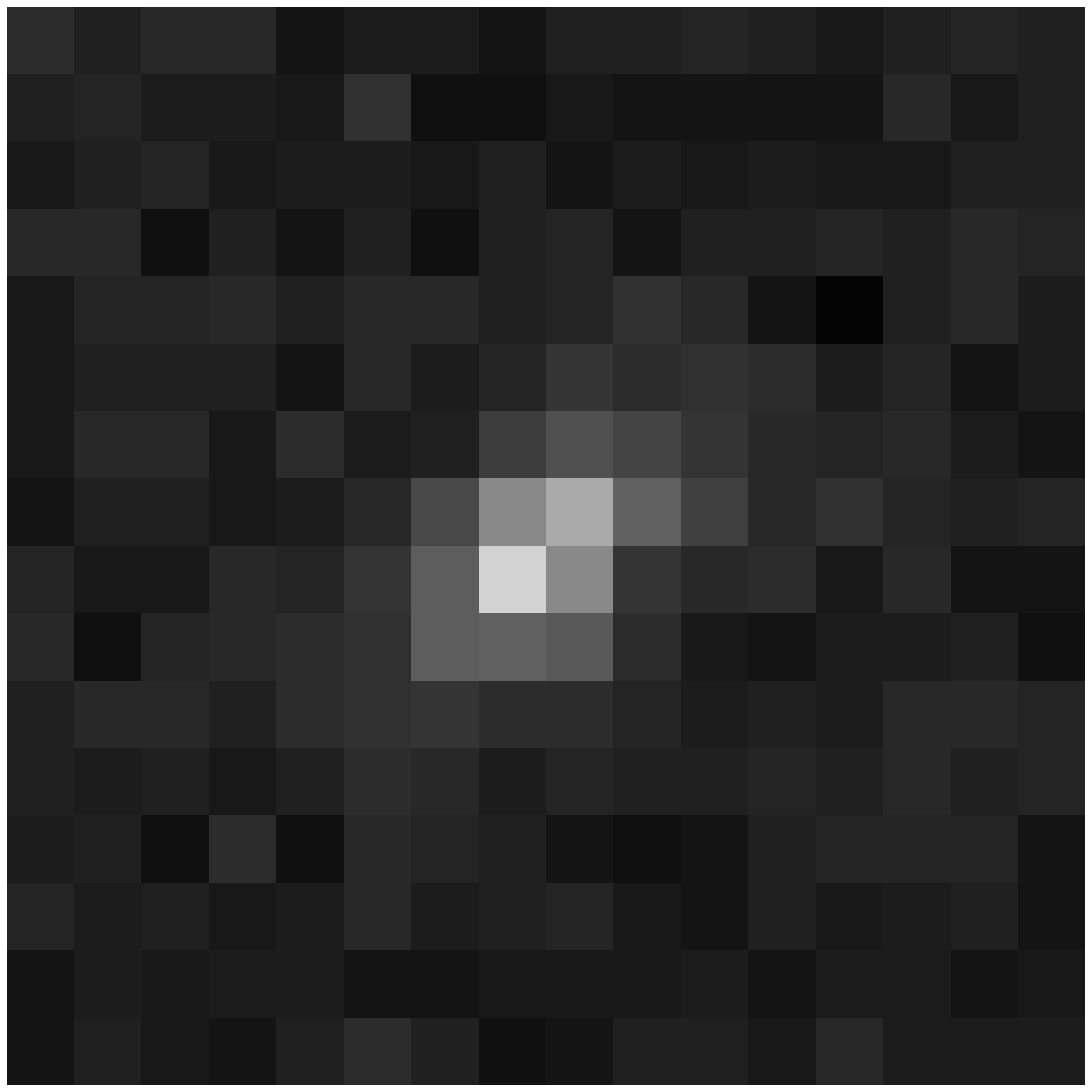} & \includegraphics[width=0.9in]{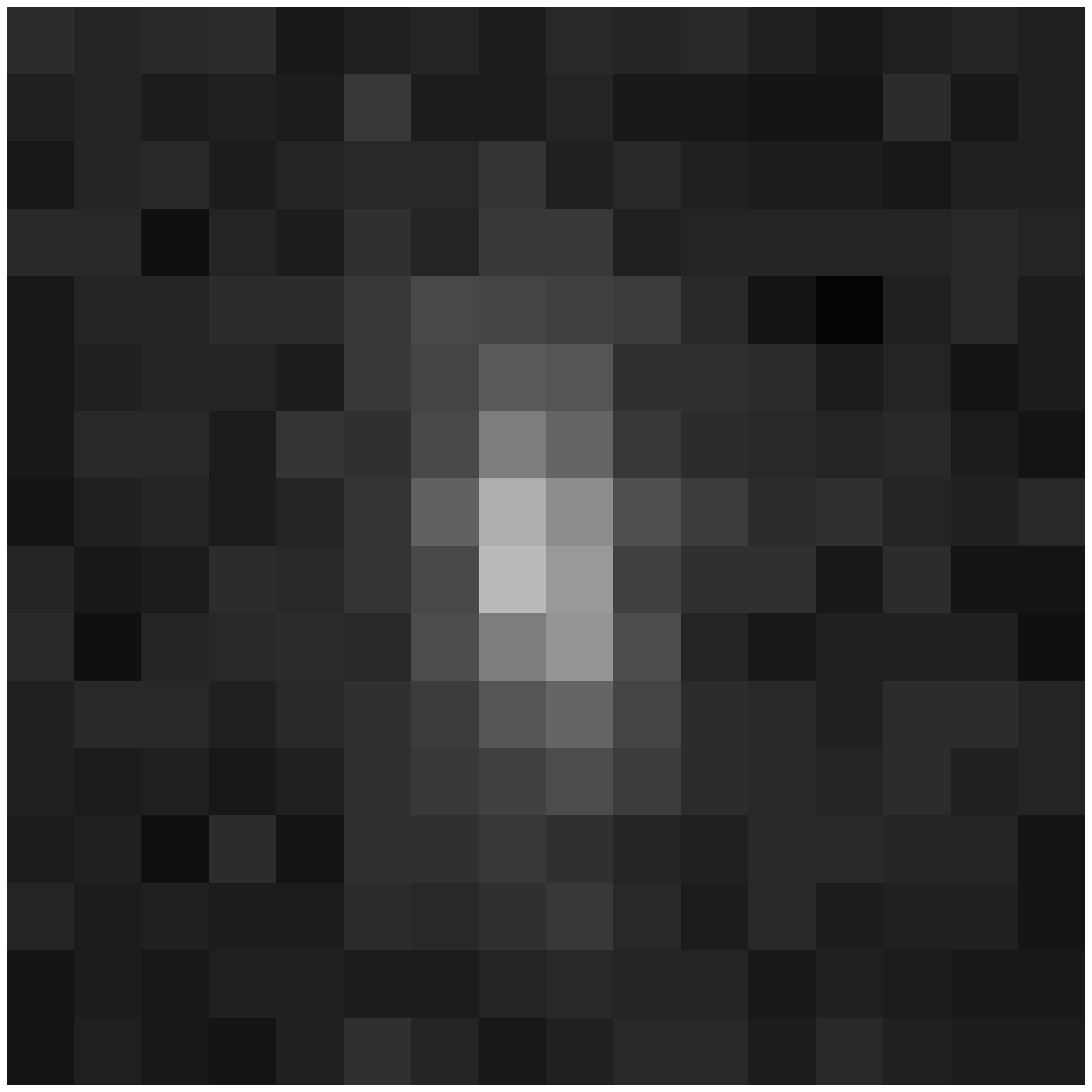}\\
	exponential & exponential & generic \sersic\\
	\end{tabular}
	\caption{A subset of noisy images for our experiment. Top: Circular profiles to test the full fitting set.
Bottom: Elliptical profiles to test the reduced fitting set. De Vaucouleurs and exponential profiles have $\lambda=4$ and $\lambda=1$, respectively.}
	\label{fig:source_images}
\end{figure}

Our experiment at this stage is not a rigorous test for shape measurement, as our test cases (\fig{source_images}) are
idealized profiles without PSF convolution. Rather, we are exploring the effect of using a variety of weight
functions with different ($k$, $\lambda$). Our test cases consist of both circular and elliptical profiles.
We generated
two-dimensional reduced $\chi^2$ maps of $k$ vs order at fixed $\lambda$ values. 

The $\chi_{\rm red}^2$ maps are then compared against those generated using Polar Shapelets, which we
will refer to simply as ``Shapelets'' in the following discussion.
In Shapelets' case, the $r$ coordinate is also normalized to $0<r<1$. The ``scaling-factor''
$\beta$ in Shapelets is now comparable to $k$ in Sersiclets, which is relative to the size of
the image. The model fits for both Sersiclets and Shapelets were computed using
{\sc Hrothgar}\footnote{http://hrothgar.sourceforge.net/} implemented in C.



\subsection{Results -- Circular models using the full fitting set}
\begin{figure}
	\begin{tabular}{cc}
		Sersiclets & Shapelets \\
		\includegraphics[width=1.2in]{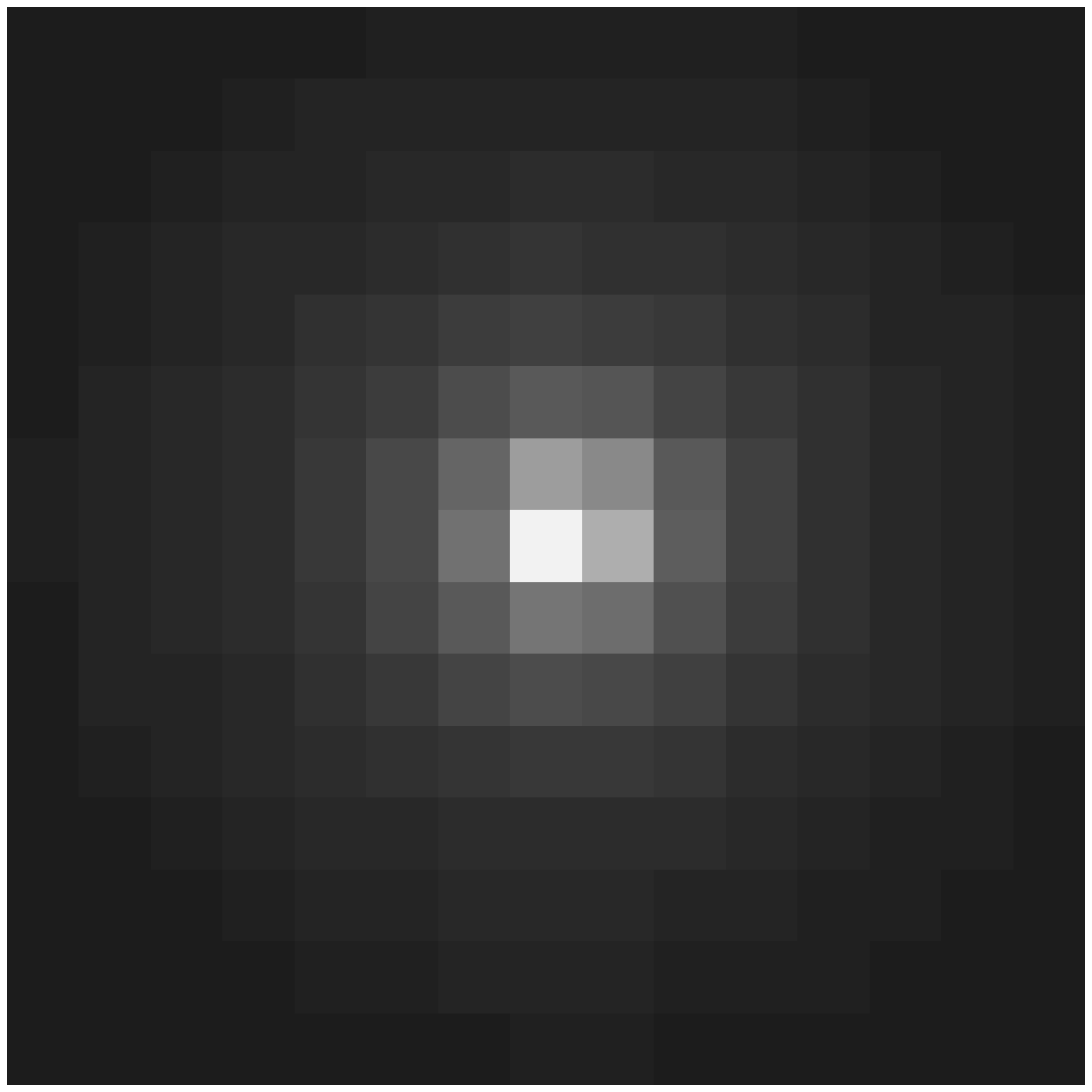} & \includegraphics[width=1.2in]{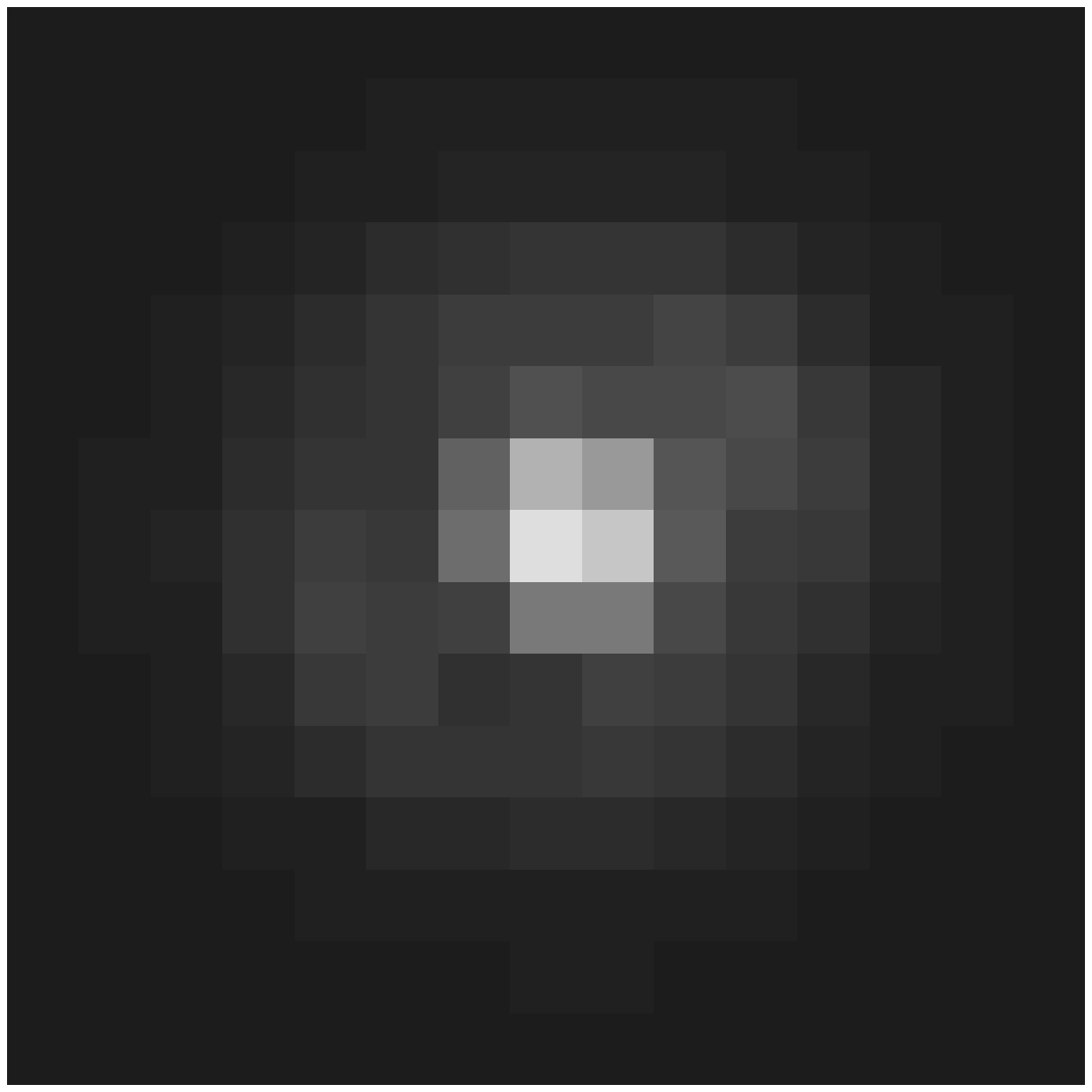}\\
		\includegraphics[width=1.6in]{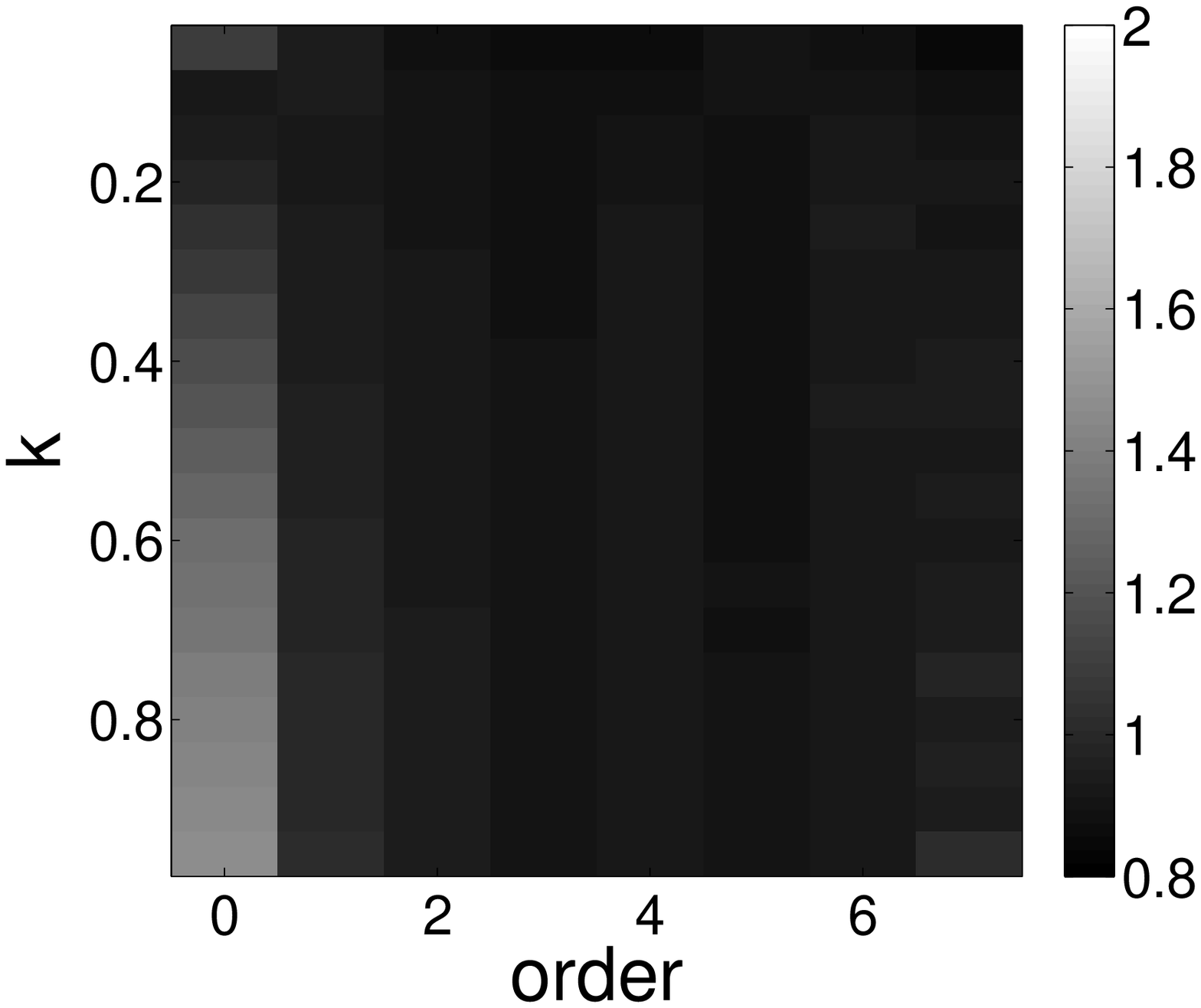} & \includegraphics[width=1.6in]{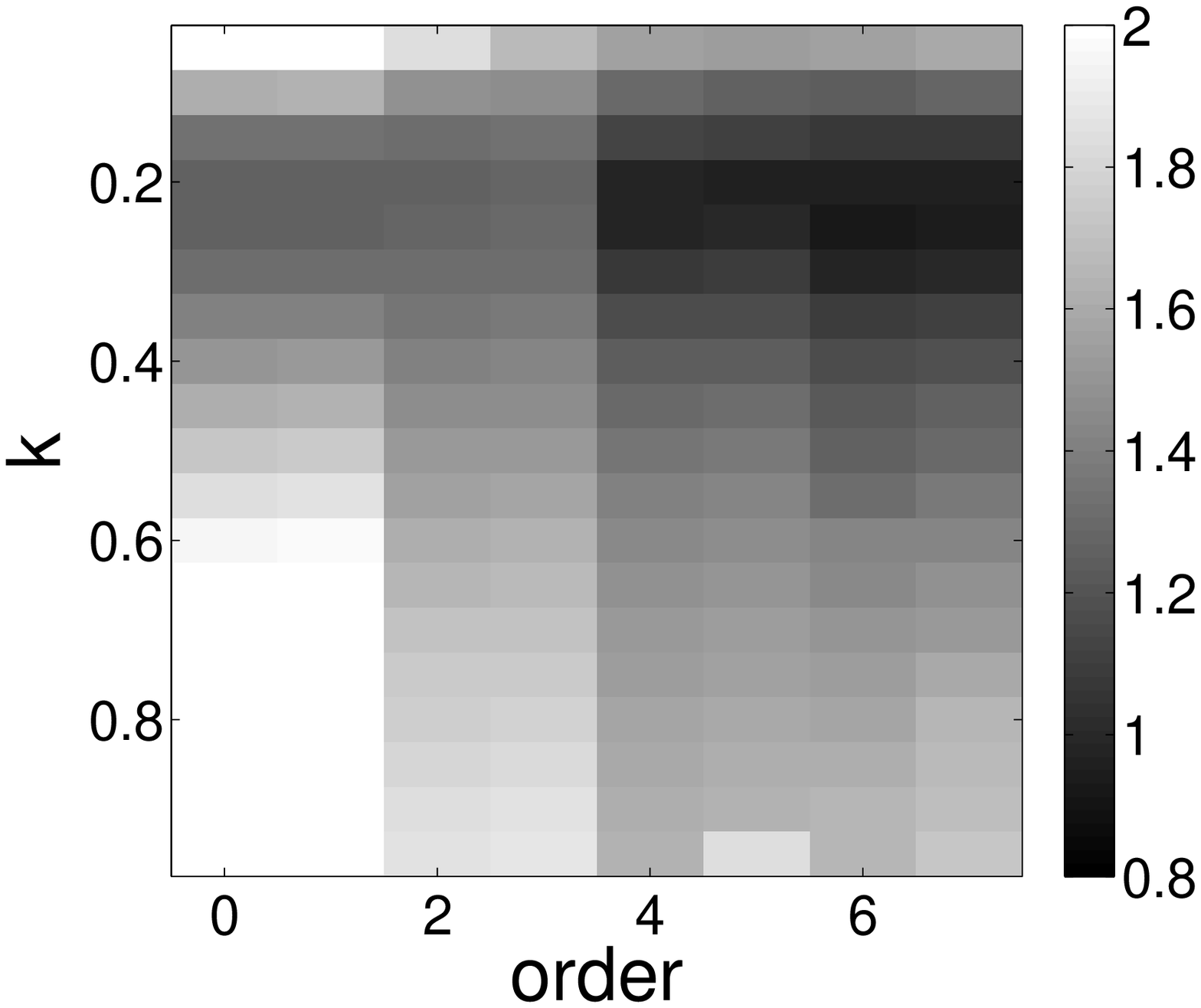}\\
	\end{tabular}
	\caption{Model fits to the de Vaucouleurs ($1/\lambda=0.25$) profile using the full fitting set. Upper panels: Sersiclets (order 1, $1/\lambda=0.3, k=0.3$) (left) and Shapelets
(order 4, $k=\beta=0.2$) (right). $\chi_{\rm red}^2$ values are $0.9371$ and $0.9737$, respectively.
Lower panels: $\chi_{\rm red}^2$ maps of the same fits as the upper panels.}
	\label{fig:circular_result}
\end{figure}

We first test the full fitting set by fitting against circular profiles. In \fig{circular_result}, we see that convergence
of $\chi_{\rm red}^2 \approx 1$ is achieved very quickly. This is not surprising as the weight functions in the models
are indeed realistic. More importantly, we find that the fits are insensitive to
the choice of $(k, \lambda)$ after the first few orders. This robustness allows us to obtain good fits without searching
for an optimal $(k, \lambda)$. This result is useful in fitting large
collections of objects, where families of objects can simply share the same pair of $(k, \lambda)$
without compromise.

Comparing the $\chi_{\rm red}^2$ maps and image reconstructions between Sersiclets and Shapelets
reveals Sersiclets' advantage. In very non-Gaussian cases such as the de Vaucouleurs case,
Sersiclets converge at a much lower order than Shapelets. In fact, Shapelets require low
signal-to-noise ratios in order to render an illusion of ``good fit''. From \fig{circular_result},
we see that Shapelets' lowest order best fit at $\rmn{order}=4$ already shows
signs of noise fitting; the model is not smooth, and it shows non-circular isophotes.
Sersiclets, though, can recover the smooth and circular profile at
orders 0 or 1.

\subsection{Results -- Elliptical model using the reduced fitting set}

\begin{figure}
	\centering
	\begin{tabular}{cc}
		Sersiclets & Shapelets \\
		\includegraphics[width=1.2in]{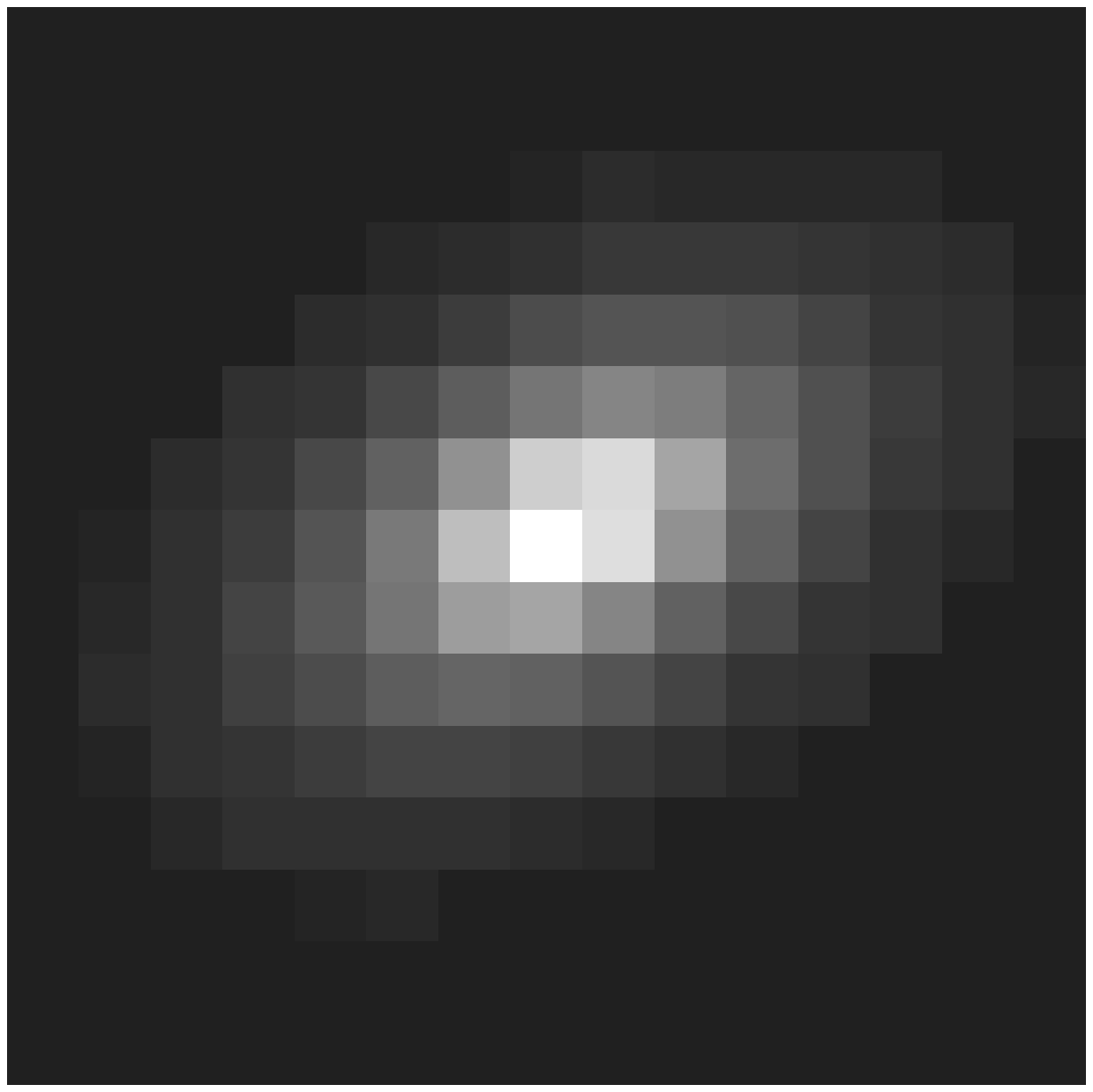} & \includegraphics[width=1.2in]{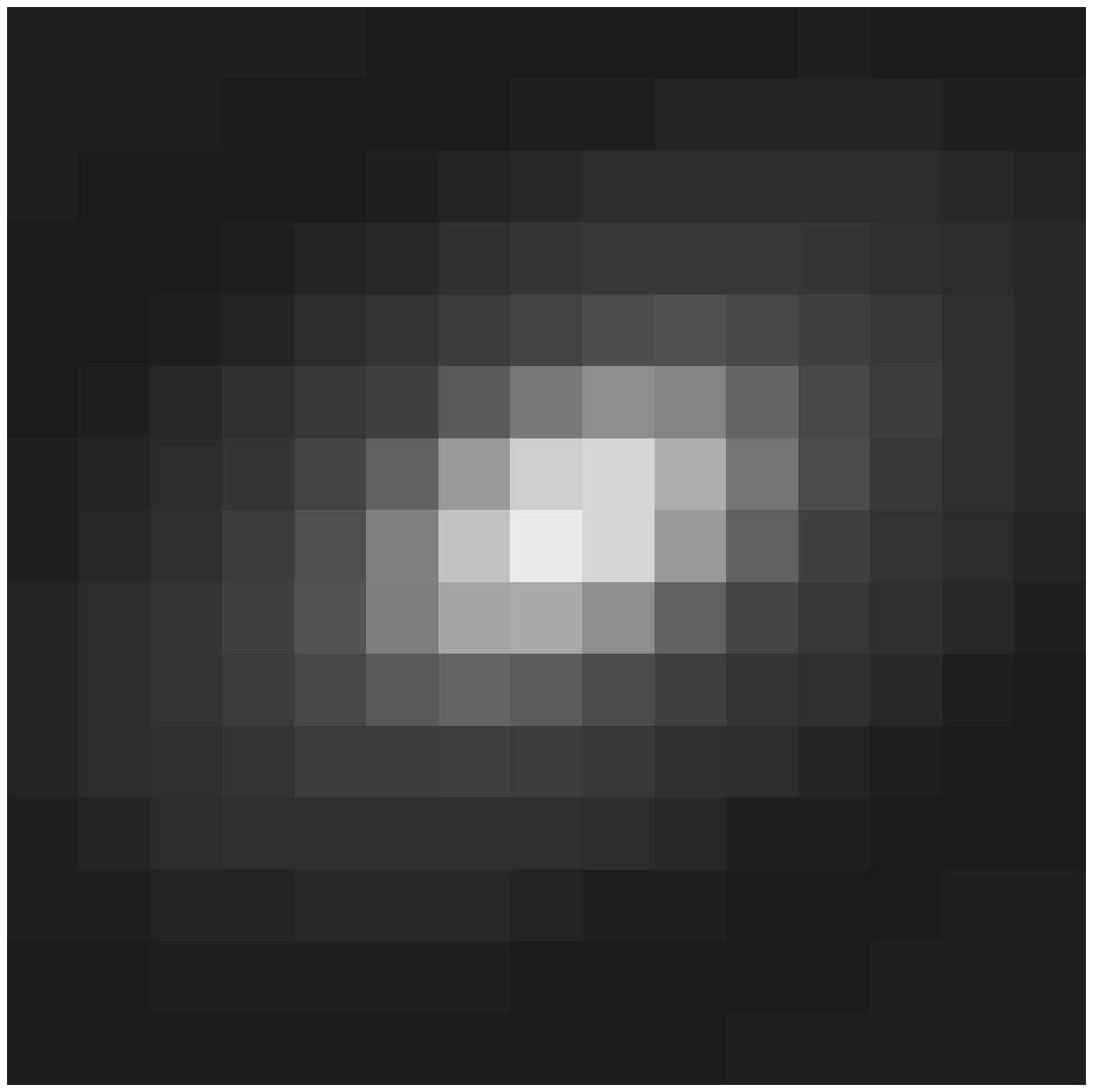} \\
		\includegraphics[width=1.6in]{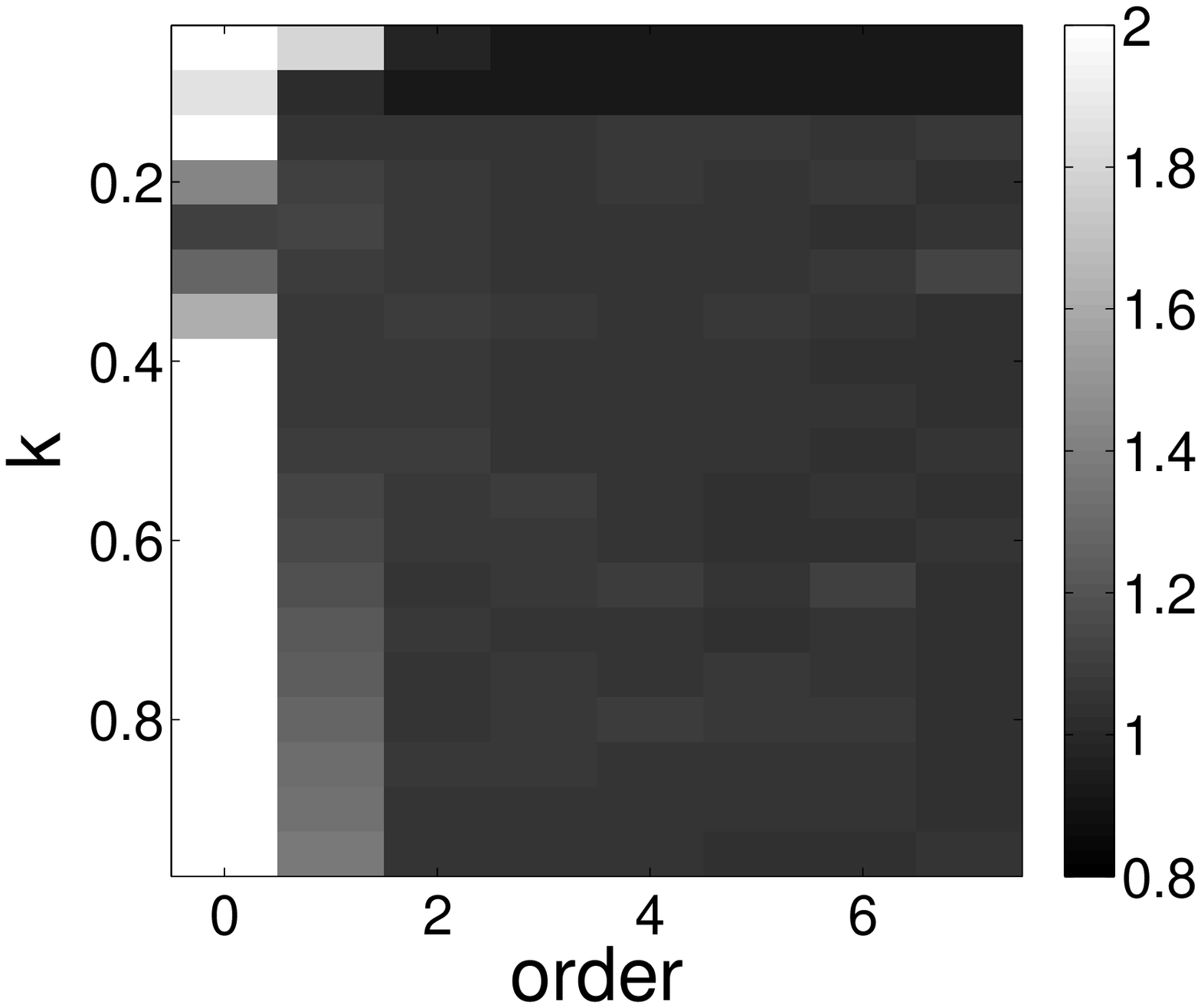} & \includegraphics[width=1.6in]{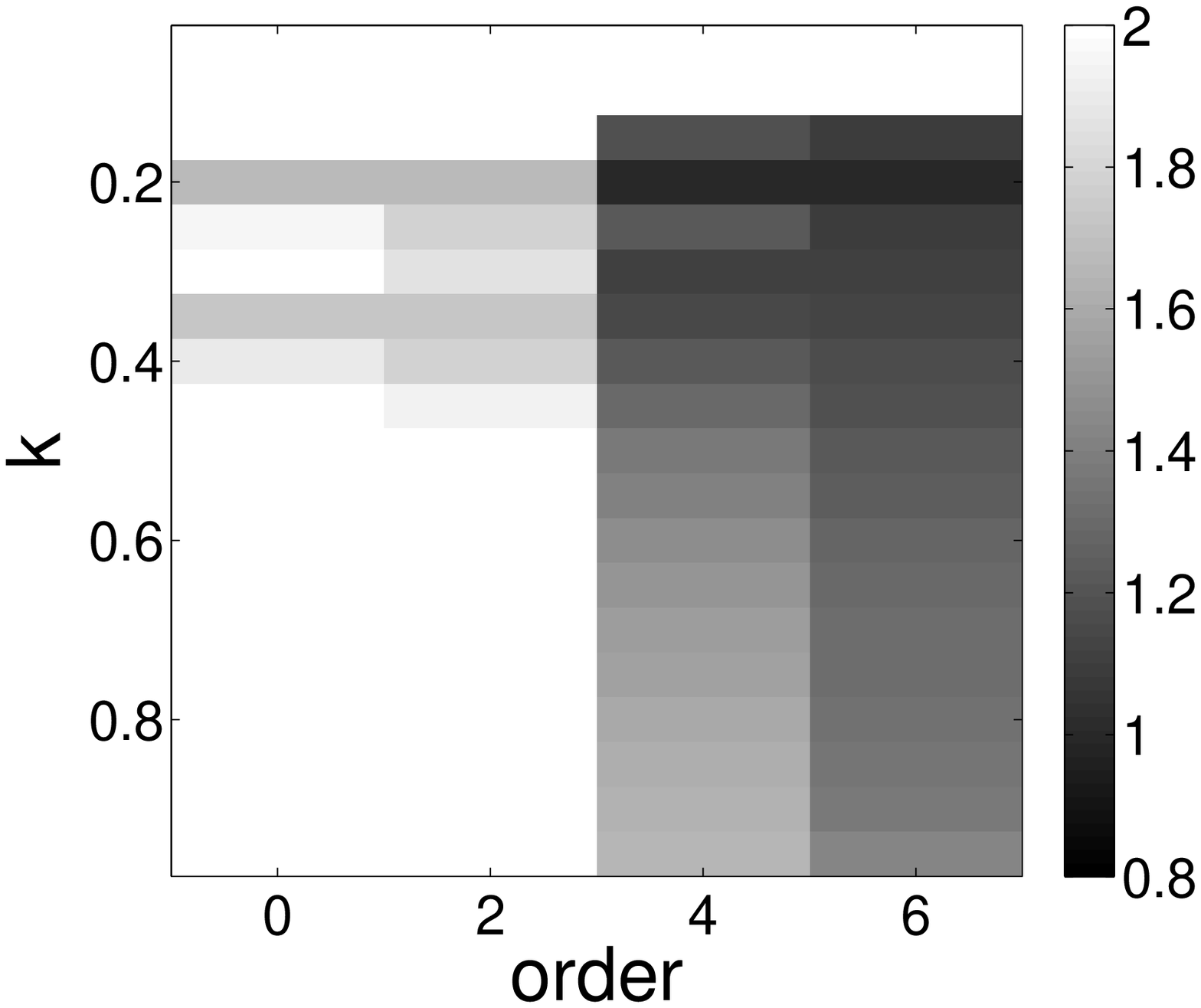}
	\end{tabular}
	\caption{Model fits to the elliptical exponential ($1/\lambda=1$) profile using the reduced fitting set. Upper panels: Sersiclets
(order 4, $1/\lambda=0.8$, $k=0.2$) (left) and Shapelets (order 4, $k=\beta=0.2$) (right). $\chi_{\rm red}^2$
values are 1.0000 and 1.0634, respectively. Lower panels: $\chi_{\rm red}^2$ maps of the same fits as the upper panels.}
	\label{fig:elliptical_result}
\end{figure}

For the elliptical profiles, the $\chi_{\rm red}^2$ maps
(\fig{elliptical_result}) are very similar to those shown in
\fig{circular_result}. This means that the $m\neq0$ components were
indeed not important, and Sersiclets' robustness in $(k,\lambda)$
are preserved. Fitting set reduction as described in \sect{basis_reduction} has been
done to both Sersiclets and Shapelets. In Shapelets' case, since
$m=0$ components do not exist for odd orders, only even orders
were possible. As seen in the image reconstruction in
\fig{elliptical_result}, the Shapelet fit no longer shows noise
fitting thanks to fitting set reduction.

\begin{figure}
	\includegraphics[width=3.3in]{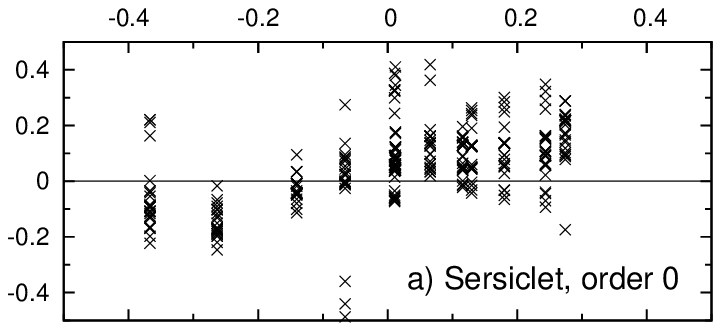}
	\includegraphics[width=3.3in]{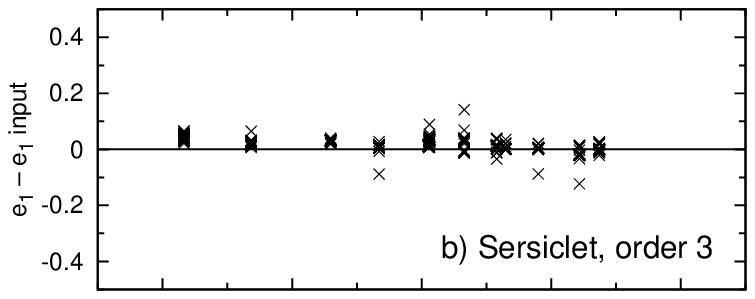}
	\includegraphics[width=3.3in]{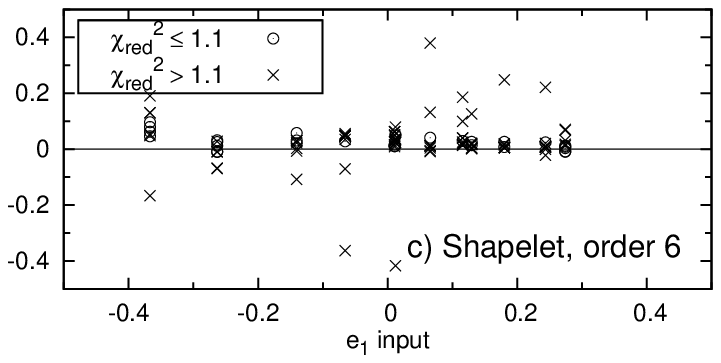}
	\caption{$e_1$ residuals measured using the reduced fitting set. Each
column of points is an object (\tab{ellipticals}), and each data point is a different
$(k,\lambda)$, where $0.05\le k \le 0.95$ in steps of 0.1 and $0.4\le 1/\lambda \le 1.0$ in steps of 0.2 (for Sersiclets). The behavior of the best fit $e_2$ is very similar
to those in $e_1$. At order 3, Sersiclets can achieve $\chi_{\rm red}^2\approx1$ and measure
$\bmath{e}$ with only a small dispersion. In contrast, Shapelets can do so only at a certain range of
$k$ (or $\beta$) that produces good fits (See \fig{elliptical_result}). At a given ``order'', Sersiclets have about twice as many degrees
of freedom as Shapelets.}
	\label{fig:e1_result}
\end{figure}

The residuals of measured $e_1$ values corresponding to each input
value have been plotted in \fig{e1_result}. It is clear that as the
fit improves with higher orders, the scatter in the measured
ellipticity is reduced. For Shapelets, the solution does
not show robustness in $k$ as the measured ellipticities are more
scattered than the Sersiclets' case.



\begin{table}
	\caption{Input parameters for the elliptical models in \fig{e1_result}. Each model is $16 \times 16$ pixels in dimension, and has peak $S/N \approx 25$.}
	\label{table:ellipticals}
	\begin{tabular} {rrrr}
		\hline
		$e_1$		&	$e_2$		&	$k$	&	$1/\lambda$\\
		\hline
		0.130		&	0.225		&	0.43	&	1.0 \\
		0.244		&	0.089		&	0.43	&	1.0 \\
		0.180		&	0.186		&	0.43	&	1.0 \\
		$-0.263$	&	0.164 		&	0.24	&	1.0 \\
		0.274		&	0.146		&	0.24	&	1.0 \\
		0.011		&	0.310		&	0.24	&	1.0 \\
		0.012		&	0.069		&	0.29	&	0.75 \\
		0.066		&	0.024		&	0.29	&	0.75 \\	
		$-0.066$	&	0.024		&	0.29	&	0.75 \\
		$-0.140$	&	0.348		&	0.28	&	0.75 \\
		0.116		&	0.357		&	0.28	&	0.75 \\
		$-0.367$	&	$-0.078$	&	0.28	&	0.75 
	\end{tabular}
\end{table}


\subsection{Discussion}

Sersiclets' robustness in $(k, \lambda)$ and its ability to converge
in relatively low orders comes from the abundance of its degrees of
freedom in the model. For the full fitting set, \eqn{linear_combo} has
$(n_\rmn{max}+1)^2$ terms in the summation as the angular quantum
number $m$ increases in steps of 1. Polar Shapelets, however, have only
$(n_\rmn{max}+1)(n_\rmn{max}+2)/2$ terms as $m$ increases
in steps of 2 at a given order of $n$. For the reduced fitting set, Sersiclets have $n_\rmn{max}+1$
terms, and Polar Shapelets have only $n_\rmn{max}/2+1$ terms.
Together with $\lambda$ and $k$, Sersiclets would have about twice
as many degrees of freedom as Polar Shapelets. To facilitate a fair
comparison, \fig{e1_result} compares Sersiclets at order 3 against
Shapelets at order 6 in the second and third panels. This ensures that 
the test uses the same degrees of polynomials available to each decomposition
technique.


A drawback of Sersiclets is numerical instability. Hermite or
associated Laguerre polynomials in Shapelets can be generated very
easily with elementary operations, whereas the general Sersiclet
polynomials require gamma functions. In particular,
the analytical solution to the integral
\begin{equation}
	\int_0^1 r^j \exp[-b_\lambda(r/k)^{1/\lambda}] dr \qquad (j=1,2...)
	\label{eqn:num_int}
\end{equation}
which is ubiquitous in \sect{radial_component}, can be written as
\begin{equation}
	\lambda(b_\lambda k^{-1/\lambda})^{-\lambda-j\lambda} \left[ \Gamma(\lambda+j\lambda) - \Gamma(\lambda+j\lambda, b_\lambda k^{-1/\lambda}) \right]
\label{eqn:num_int_soln}
\end{equation}
where $\Gamma(z)$ and $\Gamma(a,z)$ are the complete and incomplete gamma functions,
respectively, and $b_\lambda = 2\lambda-1/3$ as before. At first glance, it may
be tempting to use the analytical solution because it exists and gamma functions can
be readily evaluated. Upon closer inspection, though,
\eqn{num_int_soln} evaluates the difference between two gamma functions, each on the order
of $(\lambda+j\lambda)!$. The difference is then scaled by a number raised
to the power of $-(\lambda+j\lambda)$. One needs to be careful when evaluating such
an expression if $(\lambda+j\lambda)!$ becomes large, especially when using
numerical libraries with single precision. Alternatively, the integrand in \eqn{num_int} is a smooth
function, so the integral can be evaluated directly without implementing the analytical solution.



Sersiclets can be generalized to
different weight functions. In our derivation, although we have
chosen the \sersic\ function as our weight function, the process would
still be the same for any weight functions which are intrinsically
elliptical without explicit angular dependence. This allows for
modeling different types of objects such as the PSF using the Moffat
profile as the weight function. The integration limit of the radial
component, as we discussed in \sect{integration_limits}, can be
either finite or infinite depending on whether the weight function
is sufficiently localized.


\section{Conclusion}
\label{sec:conclusion}
We presented an extension of Shapelets by using an arbitrary
weight function in place of the Gaussian function. As galaxies'
light profiles follow the \sersic\ profile on average, we used the
\sersic\ function as our weight function. This allowed us to fit cuspy
galaxies at lower orders than Shapelets could.

Because the \sersic\ function lacks analytical properties, we used the
Gram-Schmidt process to generate the orthonormal polynomials as
radial components for the fitting functions, where the
integrals in the process must be evaluated numerically. As the
\sersic\ profile has poor local support, the integration limit must be truncated
to a finite limit.

We found that the full set of fitting functions for Sersiclets cannot
decompose an arbitrary image even at high orders, as the fitting functions
do not form a complete set. Instead of
modeling objects using all fitting functions, we reduced
the fitting set to only the circularly symmetric components ($m=0$). The
model was then sheared by $e_1$ and $e_2$ to render elliptical
shapes. The reduced set of fitting functions defines a hybrid method
which combines the most interesting features of the basis
decomposition and the fitting technique.

Our experiments so far only
focused on idealized images simulated using known profiles and
noise. Both the full and the reduced Sersiclets outperformed
Shapelets, as we expected. The Shapelet matched filter's true
performance will be tested in a
future paper on image simulations such as those for GREAT08
\citep{Br2008}. The C code to evaluate Sersiclet models is publicly
available on request.


\section*{Acknowledgments}

We are grateful to Ariel Zhitnitsky for inspiring discussions at the
beginning of this work, and Peter Melchior for valuable comments. We
also thank the anonymous referee for detailed and very constructive
comments on the manuscript.
Wayne Ngan, LVW, and Henk Hoekstra are supported by NSERC and CIfAR.

\bsp

\label{lastpage}


\begin{thebibliography}{99}
\bibitem[\protect\citeauthoryear{Berg\'e et al.}{2008}]{BPR2008}
Berg\'e, J., Pacaud, F., et al., 2008, MNRAS, 385, 695.
\bibitem[\protect\citeauthoryear{Bernstein \& Jarvis}{2002}]{bj2002}
Bernstein, G., Jarvis, M., 2002, AJ, 123, 583.
\bibitem[\protect\citeauthoryear{Bridle et al.}{2008}]{Br2008}
Bridle, S., et al., 2008, arXiv:0802.1214.
\bibitem[\protect\citeauthoryear{Heymans et al.}{2006}]{CH2006}
Heymans, C., Van Waerbeke, L., et al., 2006, MNRAS, 368, 1323.
\bibitem[\protect\citeauthoryear{Hoekstra \& Jain}{2008}]{HJ2008}
Hoekstra, H., Jain, B., 2008, arXiv:0805.0139.
\bibitem[\protect\citeauthoryear{Kitching et al.}{2008}]{K2008}
Kitching, T., Miller, L., et al. 2008, arXiv:0802.1528.
\bibitem[\protect\citeauthoryear{Kuijken}{2006}]{KK2006} Kuijken,
K., 2006, A\& A, 456, 827.
\bibitem[\protect\citeauthoryear{Massey et al.}{2007}]{M2007}
Massey, R., Heymans, C., et al., 2007, MNRAS, 376, 13.
\bibitem[\protect\citeauthoryear{Massey \& Refregier}{2005}]{mr1}
Massey, R., Refregier, A., 2005, MNRAS, 359, 1277.
\bibitem[\protect\citeauthoryear{Miller et al.}{2007}]{MKH2007}
Miller, L., Kitching, T., et al., 2007, MNRAS 382, 315.
\bibitem[\protect\citeauthoryear{Munshi et al.}{2008}]{munshi2008}
Munshi, D., Valageas, P., et al., 2008, PhR,
462, 67.
\bibitem[\protect\citeauthoryear{Peng et al.}{2002}]{peng2002}
Peng, C., Ho, L., et al., AJ, 2002, 124, 266.
\bibitem[\protect\citeauthoryear{Refregier}{2003}]{ref1} Refregier, A.,
2003, MNRAS, 338, 35.
\bibitem[\protect\citeauthoryear{Refregier \& Bacon}{2003}]{rb2003} Refregier,
A., Bacon, D., 2003, MNRAS, 338, 48.
\bibitem[\protect\citeauthoryear{S\'ersic}{1968}]{sersic1968} S\'ersic J. L., 1968, Atlas de Galaxias Australes (C\'ordoba: Obs. Astron., Univ. Nac. C\'ordoba).

\end{thebibliography}
\end{document}